\documentclass[prd,aps,nofootinbib,onecolumn,showpacs,12pt]{revtex4-1}
\usepackage{graphicx,epsfig}

\usepackage{epsf}
\usepackage{graphicx}

\begin{document}

\title{  \bf  Spectrum of  the heavy axial vector $\chi_{b1}(1P)$ and
$\chi_{c1}(1P)$ mesons in thermal QCD }
\author{ E. Veli Veliev$^{*1}$,
K. Azizi$^{\dag2}$, H. Sundu$^{*3}$, G. Kaya$^{*4}$ \\
$^{*}$Department of Physics, Kocaeli University, 41380 Izmit,
Turkey\\
 $^{\dag}$Department of Physics,
Do\u gu\c s University,
 Ac{\i}badem-Kad{\i}k\"oy, \\ 34722 Istanbul, Turkey\\
 $^1$ e-mail:elsen@kocaeli.edu.tr\\
$^2$e-mail:kazizi@dogus.edu.tr\\
$^3$email:hayriye.sundu@kocaeli.edu.tr\\
$^4$email:gulsahbozkir@kocaeli.edu.tr }

\begin{abstract}
Using the additional operators
coming up at finite temperature, we calculate the masses and
 decay constants of the $P$ wave heavy axial-vector $\chi_{b1}$ and
$\chi_{c1}$ quarkonia in the framework of  thermal QCD sum rules. In
the calculations, we take into account the perturbative two loop order
$\alpha_{s}$ corrections and nonperturbative effects up to the
dimension four condensates. It is observed that the masses
and  decay constants almost remain unchanged with respect to
the variation of the temperature up to $T\simeq 100 ~MeV$, however
after this point, the decay constants  decrease sharply and approach  approximately to zero at critical
temperature. The decreasing in  values of  the masses is also considerable after  $T\simeq 100 ~MeV$.
\end{abstract}

\maketitle


\section{Introduction }

Investigation of the in medium properties of heavy mesons such
as bottomonium ($\bar b b$) and charmonium ($\bar c c$) are of
considerable interest for  hadron physics to date. These
quarkonia  play an important role in obtaining information
on the restoration of the spontaneously broken chiral symmetry in a
nuclear medium and understanding quark gluon plasma (QGP) as
a new phase of hadronic matter. Such investigations can also provide us
with substantial knowledge on the nonperturbative QCD and interaction of quarks and gluons with QCD vacuum.

 In the last twenty years, a lot
of theoretical and experimental works have been devoted to study the
behavior of the heavy mesons in  medium. The $J/\psi$ suppression which can be considered as an indication of QGP \cite{Matsui} has been observed
in heavy ion collisions  in super proton synchrotron (SPS) at CERN and relativistic heavy ion collider (RHIC) at BNL. A plenty of theoretical works have also been dedicated to study the thermal behavior of
hadronic parameters as well as QCD degrees of freedom (for some of them and discussion on the QGP phase see for instance
 \cite{K.Morita,O.Kaczmarek,Loewe,Cheng,Miller,Nora,Nora1,Aoki,Aoki1,Borsanyi,Borsanyi1,Aoki2,S.Mallik Mukherjee,S.Mallik sarkar,E.V.Veliev,C.A.Dominguez,C.A.Dominguez2,E.V.Veliev2,F. Klingl,K.Morita2,E.V.Veliev3,E.V.Veliev4,E.V.Veliev5,Azizi}).

 Hadrons are formed in a region of energy very far from the perturbative region, hence to calculate their parameters we need to have some nonperturbative approaches. The QCD sum rules as one of the most attractive,
applicable and powerful techniques has been in the focus of much attention during last 32 years. This approach at zero temperature proposed in \cite{Shifman} and have applied to many decay channels in this period
giving results in a good consistency with the existing experimental data as well as lattice QCD calculations. This method then was extended to finite temperature QCD in \cite{Bochkarev}.
 There are two new aspects
in this extension compared to the case at zero temperature \cite{E.V.Shuryak,T.Hatsuda,S.Mallik}, namely  interaction of the particles  in the medium with the currents requiring
modification of the hadronic spectral density as well as the breakdown of the Lorentz invariance via the choice of reference frames. Because of residual
$O(3)$ symmetry at finite temperature, more operators with the same dimensions come out in the operator product expansion (OPE) compared to that of  vacuum.

The purpose of this paper is to calculate the masses and decay
constants of the $P$ wave heavy axial vector $\chi_{b1}$ and
$\chi_{c1}$ mesons in the framework of the thermal QCD sum rules. In
our calculations, we use thermal propagator containing new
non-perturbative contributions appearing at finite temperature, and
take into account the perturbative two-loop order corrections to the
correlation function \cite{Shifman,L.J.Reinders}. We use the
expressions of the temperature-dependent energy-momentum tensor
obtained via Chiral perturbation theory \cite{P.Gerber} and lattice
QCD \cite{Cheng,Miller} as well as temperature-dependent gluon
condensates and continuum threshold to obtain the behavior of the
masses and decay constants of these mesons in terms of temperature.

\section{Thermal QCD  Sum Rule for $P$ Wave Heavy Axial Vector quarkonia }
In order to extract the sum rules for the masses and  decay constants of the heavy axial vector
 $\chi_{b1}$ and $\chi_{c1}$ mesons at finite temperature, we start considering the following
two-point thermal correlation function:
\begin{eqnarray}\label{Eq1}
\Pi_{\mu\nu}\Big(q,T\Big)=i \int d^{4}x~ e^{iq\cdot x}  \langle
{\cal T}(J_{\mu}(x)J^{\dag}_{\nu}(0))\rangle,
\end{eqnarray}
where, $J_{\mu}(x)=:\overline{Q}(x)\gamma_{\mu}
\gamma_{5}Q(x):$ with $Q=b$ or $c$ is the interpolating current of heavy axial vector meson, $T$ is temperature and
${\cal T}$ indicates the time ordering product. The thermal average of
any operator $O$ is defined as
\begin{equation}\label{eqn2}
\langle O\rangle=\frac{Tr ( e^{-\beta H}O)}{Tr( e^{-\beta H})},
\end{equation}
where $H$ is the QCD Hamiltonian and  $\beta=1/T$.

According to the general philosophy of the  QCD sum rules formalism,
the above correlation function can be calculated in two different
ways. Once, in terms of QCD  degrees of freedom by the help of OPE
called the theoretical or QCD side. The OPE incorporates the effects
of the QCD vacuum through an infinite series of condensates of
increasing mass dimensions. The second, in terms of  hadronic
parameters called the physical or phenomenological side. Matching
then these two representations, we find  sum rules for the physical
observables under consideration. To suppress the contribution of the
higher states and continuum, we apply Borel transformation as well
as continuum subtractions. In the following, we calculate the
correlation function in two aforesaid windows.

\subsection{The phenomenological side}

Technically, to obtain the physical or phenomenological side of the correlation
function, we  insert a complete set of intermediate hadronic
states with the same quantum numbers as the interpolating current
$J_{\mu}(x)$ into the correlation function.  After performing the four-integral over $x$ and isolating the ground state contribution, we get
\begin{eqnarray}\label{phe}
\Pi_{\mu\nu}(q)=\frac{f_{A}^2M_{A}^2}{M_{A}^2-q^2}\left(-g_{\mu\nu}
+\frac{q_\mu q_\nu}{q^2}\right) +\cdots,
\end{eqnarray}
where the  $f_{A}$ and $M_{A}$ are decay constant and mass
of the heavy axial vector meson, respectively. The  $.....$ in the above equation stands for the
contribution of the  excited heavy axial vector states and continuum.
In deriving the Eq. (\ref{phe}), we have defined the decay constant $f_{A}$ by the matrix element of the
current $J_{\mu}$ between the vacuum and the  mesonic state in the following manner:
\begin{eqnarray}\label{Jmu}
{\langle}0|J_{\mu}|A(q,\lambda){\rangle}=f_{A} m_{A}
\varepsilon^{(\lambda)}_{\mu},
\end{eqnarray}
where  $\varepsilon_\mu$ is the  four-polarization vector. We have also used the summation over polarization vectors as
\begin{eqnarray}\label{sum}
\sum_{\lambda}\varepsilon^{(\lambda)^*}_{\mu}\varepsilon^{(\lambda)}_{\nu}=-(g^{\mu\nu}-q_{\mu}q_{\nu}/m_{A}^2).
\end{eqnarray}

\subsection{The QCD side}

In QCD side, the correlation function is calculated in deep
Euclidean region where $q^2\ll-\Lambda_{QCD}^2$ via OPE where the short
or perturbative and long distance or non-perturbative effects are
separated, i.e.,
\begin{eqnarray}\label{correl.func.QCD1}
\Pi_{\mu\nu}^{QCD}(q,T)
=\Pi_{\mu\nu}^{p}(q,T)+\Pi_{\mu\nu}^{np}(q,T).
\end{eqnarray}
The short distance contributions are calculated using the
perturbation theory, while the long distance contributions
 are expressed in terms of the thermal expectation
values of the quark and gluon condensates as well as thermal average
of the energy density coming up at finite temperature.

In the rest frame of the medium for axial vector meson at
rest, the  correlation function in QCD side can be written in terms of the transverse and longitudinal components as
\begin{equation}\label{eq:proj_VA}
\Pi_{\mu\nu}(q)=\left(\frac{q_\mu q_\nu}{q^2}-g_{\mu\nu}\right)
\Pi_t(q)+\frac{q_\mu q_\nu}{q^2} \Pi_l(q),
\end{equation}
where the functions, $\Pi_t(q)$ and $ \Pi_l(q)$ are found in terms of the total correlation function as
\begin{eqnarray}
 \Pi_t(q)&=&\frac{1}{3} \left( \frac{q^\mu q^\nu}{q^2} -
g^{\mu\nu} \right)
 \Pi_{\mu\nu}(q),\nonumber\\ \Pi_l(q) &=& \frac{1}{q^2}
q^\mu q^\nu \Pi_{\mu\nu}(q).
\end{eqnarray}
Here, we would like to mention that the transverse and longitudinal components are related to each other, hence it is enough to use one of them to obtain the  thermal sum rules for the physical quantities
under consideration.
Here, we use the function $ \Pi_t(q)$ for this aim.  It can  be shown that this function for the fixed values of the
$|\textbf{q}|$,   can be written as  \cite{S.Mallik Mukherjee}:
\begin{eqnarray}\label{Eq4}
\Pi_{t}\Big(q_0^2,
T\Big)=\int^{\infty}_{0}{dq_0^{\prime}}^{2}~\frac{\rho_{t}\Big({q_0^{\prime}}^{2},
T\Big)}{{q_0^{\prime}}^{2}+Q_0^2},
\end{eqnarray}
 where $Q_0^2=-q_0^2$, and
\begin{eqnarray}
\rho_{t}\Big(q_0^{2}, T\Big)=\frac{1}{\pi}Im\Pi_{t}\Big(q_0^{2},
T\Big)\tanh\frac{\beta q_0}{2},
\end{eqnarray}
is the spectral density. We also should stress that the function $\Pi_{t}\Big(q_0^2,T\Big)$ receives contributions from both annihilation and scattering parts (for more information see \cite{E.V.Veliev5}). However,
as we deal with the mesons containing quark and antiquark with the same masses, the scattering part gives  zero and here we focus our attention to calculate only the annihilation part.

The thermal correlation function in QCD side is obtained from Eq. (\ref{Eq1}) contracting out all quark fields via Wick's theorem. As a result, we obtain the following expression in terms of thermal heavy quarks propagators:
\begin{eqnarray}\label{Eq5}
\Pi_{\mu\nu}\Big(q, T\Big)=i\int
\frac{d^4k}{(2\pi)^4}Tr\Big[\gamma_{\mu}S(k)\gamma_{\nu}S(k-q)\Big].
\end{eqnarray}
In real time thermal field theory, the function $\Pi_{t}\Big(q_0^2,T\Big)$ can be expressed in $2 \times 2$ matrix representation, the elements of which depend on
 only one analytic function. Hence calculation of the 11-component of such matrix is
enough to get information on the dynamics of the corresponding two-point correlation function.
The 11-component of the thermal quark
propagator $S(k)$ which is given as a sum of its vacuum
expression and a term depending on the temperature is given as
\cite{A.Das}:
\begin{eqnarray}\label{Eq6}
S(k)=(\gamma^{\mu}k_{\mu}+m)\Big(\frac{1}{k^2-m^2+i\varepsilon}+2\pi
in(|k_0|)\delta(k^2-m^2)\Big),
\end{eqnarray}
where  $n(x)=[\exp(\beta x)+1]^{-1}$ is the Fermi distribution
function and $m$ is the quark mass. Performing the integral over $k_0$ in the $\textbf{q}=0$ limit, we
get  the imaginary part of the $\Pi_t(q^2_{0},T)$ as:
\begin{eqnarray}\label{Lq0}
Im\Pi_t (q^2_{0},T)=-\int \frac{d\textbf{k}}{8\pi^{2}} \frac{4
m^{2}-3q_{0}\omega+2\omega^{2}}{\omega^{2}}\Big[1-2n(\omega)+2n^{2}(\omega)\Big]\delta(q_{o}-2\omega),
\end{eqnarray}
where $\omega=\sqrt{\textbf{k}^2+m^2}$. After standard calculations,
we get the following expression for the annihilation part of the
spectral density:
\begin{eqnarray}\label{Eq9}
\rho_{t,a}(s)=\frac{s}{4\pi^2}
[v(s)]^{3}\Big[1-2n\Big(\frac{\sqrt{s}}{2T}\Big)\Big] ,
\end{eqnarray}
where, $v(s)=\sqrt{1-4m^2/s}$.
As we previously mentioned, we take into account also the perturbative two-loop  $\alpha_{s}$ order correction to the
spectral density. At zero temperature,  it is given as
\cite{Shifman,L.J.Reinders}:
\begin{eqnarray}\label{Pi1}
\rho_{\alpha_{s}}(s)&=&\alpha_{s}
\frac{s}{3\pi^{3}}\Big[\pi v^{3}\Big(\frac{\pi}{2v}-\frac{1+v}{2}\Big(\frac{\pi}{2}-\frac{3}{\pi}\Big)\Big)\nonumber\\
&+&\Big(P^{A}(v)-P(v)\Big)\ln\frac{1+v}{1-v}+Q^{A}(v)-Q(v)\Big],
\end{eqnarray}
where we have set $v=v(s)$ and the functions $P(v)$, $Q(v)$, $P^{A}(v)$ and $Q^{A}(v)$ are given as:
\begin{eqnarray}\label{npPi}
P(v)&=&\frac{5}{4}(1+v^{2})^{2}-2,
\nonumber\\
Q(v)&=&\frac{3}{2}v(1+v^{2}),
\nonumber\\
P^{A}(v)&=&\frac{21}{32}+\frac{59}{32}v^{2}+\frac{19}{32}v^{4}-\frac{3}{32}v^{6},
\nonumber\\
Q^{A}(v)&=&-\frac{21}{16}v+\frac{30}{16}v^{3}+\frac{3}{16}v^{5}.
\end{eqnarray}
To get the thermal version of the above two-loop  $\alpha_{s}$ order correction,  we replace the strong coupling $\alpha_{s}$ by its temperature dependent lattice improved
version given in \cite{E.V.Veliev5} ( for more details see also \cite{K.Morita,O.Kaczmarek}).

Our final task  in this section is to calculate the
nonperturbative part of the thermal correlation function.
The nonperturbative part in our case  can be written   in terms of
operators up to dimension four as:
\begin{equation}\label{eqn4}
\Pi^{np}_t (q^2_{0},T)= C_{1}\langle \overline{\psi}\psi\rangle +
C_{2}\langle G^{a}_{\mu\nu} G^{a\mu\nu}\rangle + C_{3}\langle
u\Theta u\rangle. \\
\end{equation}
where, $C_{n}(q^{2})$ are   as Wilson coefficients. As we also previously mentioned, at
finite temperature the Lorentz invariance is broken by the choice of
 reference  frame and new operators appear in the Wilson
expansion above. The new four-dimension operator here is $\langle
u\Theta u\rangle$, where  $\Theta^{\mu \nu}$ is the energy momentum
tensor and $u^{\mu}$ is the four-velocity of the heat bath and it is
introduced to restore Lorentz invariance formally in the thermal
field theory. In the rest frame of the heat bath, we have  $ u^{\mu}
= (1, 0, 0, 0)$ which leads to $u^2 = 1$. Note that in our
calculations, we ignore the heavy quark condensate $\langle
\overline{\psi}\psi\rangle$ since it suppress by inverse powers of
the heavy quark mass.

 To proceed in calculation of the nonperturbative part, we use the
nonperturbative part of the quark propagator in an external gluon
field, $A^a_{\mu}(x)$ in the Fock-Schwinger gauge,
$x^{\mu}A^a_{\mu}(x)=0 $. In this gauge, the  vacuum gluon field  $A^a_{\mu}(k')$
is written in terms of gluon field strength tensor in momentum space as follows:
\begin{eqnarray}\label{Amu}
A^{a}_{\mu}(k')=-\frac{i}{2}(2 \pi)^4 G^{a}_{\rho
\mu}(0)\frac{\partial} {\partial k'_{\rho}}\delta^{(4)}(k'),
\end{eqnarray}
where $k'$ is the gluon momentum.
\begin{figure}
\begin{center}
\includegraphics{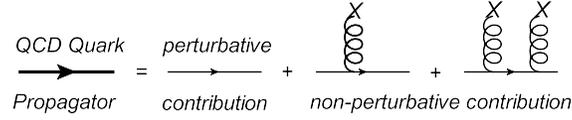}
\end{center}
\caption{\label{BRmh0}The quark propagator in the gluon background
fields.}
\end{figure}
Taking into account one and two gluon lines attached to the
quark line as shown in Fig. 1, up to terms required for our
calculations, the non-perturbative part of the temperature-dependent massive quark propagator
 is obtained as:
\begin{eqnarray}\label{Saap1}
S^{aa^{\prime }np}(k)&=& -\frac{i}{4}g (t^{c})^{aa^{\prime}}
G^{c}_{\kappa\lambda}\frac{1}{(k^2-m^2)^2}\Big[\sigma_{\kappa\lambda}
(\not\!k+m)+(\not\!k+m)\sigma_{\kappa\lambda}\Big]
\nonumber\\
&+&\frac{i~g^2~\delta^{aa^{\prime}}} {9~(k^2-m^2)^4}\Big\{\frac{3m
(k^2+m \not\!k)}{4}\langle
G^{c}_{\alpha\beta}G^{c\alpha\beta}\rangle+ \Big[ m
\Big(k^2-4(k\cdot u)^2\Big)
\nonumber\\
&+&\Big(m^2 -4(k\cdot u)^2\Big)\not\!k+4(k\cdot
u)(k^2-m^2)\not\!u\Big]\langle
u^{\alpha}\Theta^{g}_{\alpha\beta}u^{\beta}\rangle\Big\},
\end{eqnarray}
where $\Theta^{g}_{\alpha\beta}$ is the traceless gluonic part
of the energy-momentum tensor of the QCD.

Using the above expression and after straightforward but lengthy calculations, we get the following expression for the nonperturbative part:
\begin{eqnarray}\label{NonPert}
\Pi_t^{np}&=&\int_{0}^{1}~dx \Big\{\frac{g^{2}}{144
q^{2}\pi^{2}(m^{2}+q^{2}(-1+x)x)^{4}}\Big[q^{2}(9A(-8q^{6}(-1+x)^{4}x^{4}-m^{2}q^{4}(-1+x)^{2}x^{2}\nonumber \\
&\times&(-3-26x+28x^{2}-4x^{3}+2x^{4})+3m^{6}(5-16x+26x^{2}-20x^{3}+10x^{4})\nonumber \\
&+&2m^{4}q^{2}x(-6+3x+16x^{2}-33x^{3}+30x^{4}-10x^{5}))+B(-8m^{2}q^{4}(-1+x)^{2}\nonumber\\
&\times&x^{2}(4+5x-4x^{2}-2x^{3}+x^{4})-q^{6}(-1+x)^{3}x^{3}(6+19x-17x^{2}-4x^{3}+2x^{4})\nonumber\\
&-&2m^{6}(29-92x+150x^{2}-116x^{3}+58x^{4})+11m^{4}q^{2}x\nonumber\\
&\times&(6-11x+8x^{2}+x^{3}-6x^{4}+2x^{5})))-2B(-2q^{6}(-1+x)^{3}x^{3}\nonumber\\
&\times&(6-17x+19x^{2}-4x^{3}+2x^{4})-m^{2}q^{4}(-1+x)^{2}x^{2}\nonumber\\
&\times&(37-154x+188x^{2}-68x^{3}+34x^{4})+m^{6}(19-64x+102x^{2}-76x^{3}\nonumber\\
&+&38x^{4})-4m^{4}q^{2}x(-6+47x-116x^{2}+143x^{3}-102x^{4}+34x^{5}))(q\cdot
u)^{2}\Big] \Big\},
\end{eqnarray}
where $A=\frac{1}{24}\langle G^{a}_{\mu\nu}
G^{a\mu\nu}\rangle+\frac{1}{6}\langle
u^{\alpha}\Theta^{g}_{\alpha\beta}u^{\beta}\rangle$ and
$B=\frac{1}{3}\langle
u^{\alpha}\Theta^{g}_{\alpha\beta}u^{\beta}\rangle$.

\subsection{Thermal Sum Rules for Physical Quantities}
Now it is time to equate two different representations of the correlation function from physical and QCD sides and perform
 continuum subtraction  to suppress
the contribution of the higher states and continuum. As a result of this procedure we get the following sum rule including the temperature-dependent mass and decay constant:
\begin{eqnarray}\label{f2Q4}
\frac{{f_{A}^2(T)}
Q_0^4}{[m_{A}^2(T)+Q_0^2]~m_{A}^2(T)}=Q_0^4\int^{s_0(T)}_{4m^2}\frac{[\rho_{t,a}(s)+\rho_{\alpha_s}(s)]}
{s^2(s+Q_0^2)}ds+\Pi_{t}^{np},
\end{eqnarray}
where $s_0(T)$ is temperature-dependent continuum threshold and  for simplicity, the temperature-dependent width of meson has
been neglected. To further suppress the higher states and continuum contributions, we also apply the Borel transformation with respect to $Q_0^2$ to both sides of the above sum rule.
As a result we get,
\begin{eqnarray}\label{dddd}
f_{A}^2(T)m_{A}^2(T)exp (-{\frac{s}{M^2}})=\int^{s_0(T)}_{4m^2}ds
\Big[\rho_{t,a}(s)+\rho_{\alpha_s}(s)\Big]exp
(-{\frac{s}{M^2}})+\widehat{B}\Pi_t^{np},
\end{eqnarray}
where the nonperturbative part in Borel scheme is obtained as:
\begin{eqnarray}\label{BorelNonPert}
\hat{B}\Pi_t^{np}&=&\int_{0}^{1}~dx  \frac{1}{48
M^{6}\pi^{2}(-1+x)^{4}x^{4}}exp\Big[\frac{m^{2}}{M^{2}(-1+x)x}\Big]g^{2}\Big\{3A(8M^{6}(-1+x)^{4}x^{4}\nonumber\\
&+&m^{6}(1-2x)^{2}(1-2x+2x^{2})-m^{2}M^{4}(-1+x)^{2}x^{2}\nonumber\\
&\times&(-3-2x+4x^{2}-4x^{3}+2x^{4})+m^{4}M^{2}x(3-14x+14x^{2}+13x^{3}\nonumber\\
&-&24x^{4}+8x^{5}))-B(3m^{6}(1-2x)^{2}(1-2x+2x^{2})+M^{6}(-1+x)^{3}x^{3}\nonumber\\
&\times&(6-29x+31x^{2}-4x^{3}+2x^{4})-m^{2}M^{4}(-1+x)^{2}x^{2}(-4-29x\nonumber\\
&+&43x^{2}-28x^{3}+14x^{4})+m^{4}M^{2}x(8-35x+22x^{2}+69x^{3}-96x^{4}+32x^{5}))
\Big\}.
\end{eqnarray}
Here  $M^2$ is the Borel mass parameter. Considering  Eq.
(\ref{dddd}), the mass squared of the heavy axial vector meson alone
can be obtained as:
\begin{eqnarray}\label{mV2}
m^2_{A}(T)=\frac{G(T)} {F(T)},
\end{eqnarray}
where,
\begin{eqnarray}\label{ft}
F(T)=\int^{s_0(T)}_{4m^2}ds
\Big[\rho_{t,a}(s)+\rho_{\alpha_s}(s)\Big]exp
(-{\frac{s}{M^2}})+\widehat{B}\Pi_t^{np},
\end{eqnarray}
and
\begin{eqnarray}\label{gt}
G(T)=M^4 \frac{d}{dM^2}F(T).
\end{eqnarray}

\subsection{ Numerical Results}
To numerically analyze the sum rules for mass and decay constant, we use the following temperature-dependent continuum threshold   \cite{C.A.Dominguez2}:
\begin{eqnarray}\label{sT}
s_0(T)= s_{0}\left[\vphantom{\int_0^{x_2}}
1-\Big(\frac{T}{T^{*}_{c}}\Big)^8\vphantom{\int_0^{x_2}}\right]+4~m^2~
\left(\vphantom{\int_0^{x_2}}\frac{T}{T^{*}_{c}}
\vphantom{\int_0^{x_2}} \right)^8 ,
\end{eqnarray}
where $T^{*}_{c}=1.1~T_c=0.176~GeV$ with $T_c$ being critical temperature and $s_0$ is the continuum threshold at zero temperature. For the temperature-dependent  gluon condensate we also use \cite{Cheng,Miller}
\begin{eqnarray}\label{G2TLattice}
\langle G^2\rangle=\frac{\langle
0|G^2|0\rangle}{exp\left[\vphantom{\int_0^{x_2}}12\Big(\frac{T}{T_{c}}-1.05\Big)
\vphantom{\int_0^{x_2}}\right]+1} .
\end{eqnarray}
For the thermal average of total energy density $\langle \Theta \rangle$ we use both results: i) obtained in lattice QCD \cite{Cheng,Miller}:
 \begin{eqnarray}\label{tetag}
\langle \Theta \rangle= 2 \langle \Theta^{g}\rangle=
6\times10^{-6}exp[80(T-0.1)](GeV^4),
\end{eqnarray}
where
this parametrization is valid only in the region   $0.1~GeV\leq T
\leq 0.17~GeV$. ii) obtained via chiral perturbation
theory
\cite{P.Gerber}:
\begin{eqnarray}\label{tetagchiral}
\langle \Theta\rangle= \langle \Theta^{\mu}_{\mu}\rangle +3~p,
\end{eqnarray}
where $\langle \Theta^{\mu}_{\mu}\rangle$ is trace of the
total energy momentum tensor and $p$ is pressure. They
are given as:
\begin{eqnarray}\label{tetamumu}
\langle
\Theta^{\mu}_{\mu}\rangle&=&\frac{\pi^2}{270}\frac{T^{8}}{F_{\pi}^{4}}
\ln \Big(\frac{\Lambda_{p}}{T}\Big),\nonumber\\
p&=&
3T\Big(\frac{m_{\pi}~T}{2~\pi}\Big)^{\frac{3}{2}}\Big(1+\frac{15~T}{8~m_{\pi}}+\frac{105~T^{2}}{128~
m_{\pi}^{2}}\Big)exp\Big(-\frac{m_{\pi}}{T}\Big),\nonumber\\
\end{eqnarray}
where $\Lambda_{p}=0.275~GeV$, $F_{\pi}=0.093~GeV$ and
$m_{\pi}=0.14~GeV$.

 We also use the values
$m_c=(1.3\pm0.05)~GeV$, $m_b=(4.7\pm0.1)~GeV$ and ${\langle}0\mid
\frac{1}{\pi}\alpha_s G^2 \mid 0 {\rangle}=(0.012\pm0.004)~GeV^4$
for quarks
 masses and gluon condensate at zero temperature. Finally, we should find the working region for the  continuum
threshold at zero temperature ($s_0$) and Borel mass parameter ($M^2$) such that
 the physical observables are  weakly depend on  these parameters according to the standard criteria of the QCD sum rules.   The continuum threshold, $s_{0}$ is not
totally arbitrary and  it is  correlated  to the energy of the first
exited state of the heavy axial vector meson. Our numerical calculations lead to the  intervals $s_0=(106- 110)~GeV^2$ and $s_0=(15- 17)~GeV^2$ for   the  $\chi_{b1}$ and
$\chi_{c1}$ heavy axial mesons, respectively.
The working region for the Borel mass parameter is calculated
requiring  that  not only the contributions of the higher states and
continuum are efficiently suppressed but also the contributions of the operators with higher dimensions  are ignorable. We get the
working regions  $ 10~ GeV^2
\leq M^2 \leq 35~ GeV^2 $ and $ 5~ GeV^2 \leq M^2 \leq 25~ GeV^2 $
 respectively for the  $\chi_{b1}$ and $\chi_{c1}$ channels.
\begin{figure}[h!]
\begin{center}
\includegraphics[width=10cm]{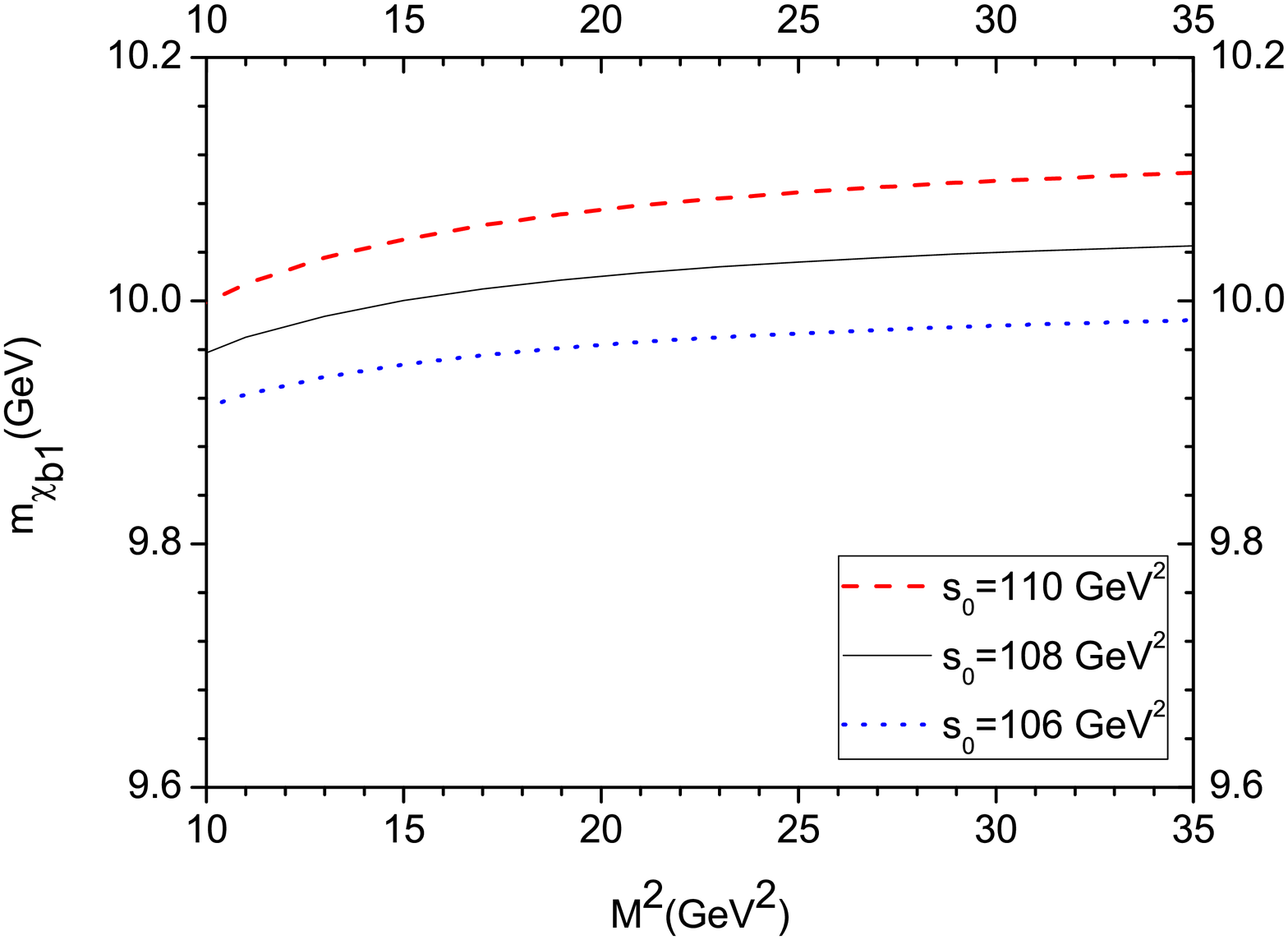}
\end{center}
\caption{Dependence of the mass of the $\chi_{b1}$  meson
on the Borel parameter $M^2$ at zero temperature.} \label{mJPsiMsq17Jan}
\end{figure}
\begin{figure}[h!]
\begin{center}
\includegraphics[width=10cm]{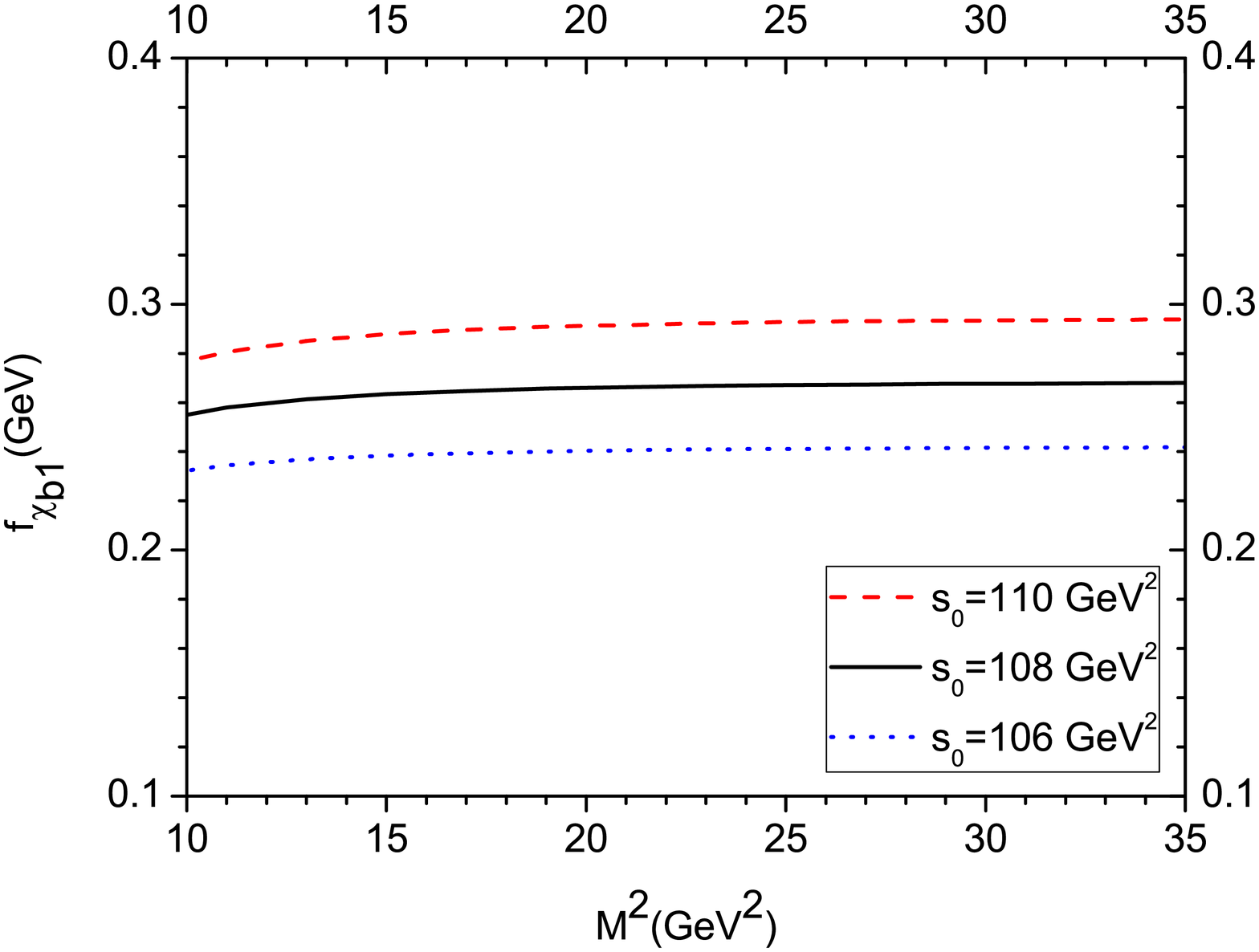}
\end{center}
\caption{Dependence of the  decay constant of the
$\chi_{b1}$ meson  on the Borel parameter $M^2$ at zero temperature.}
\label{fJPsiMsq17Jan}
\end{figure}
\begin{figure}[h!]
\begin{center}
\includegraphics[width=10cm]{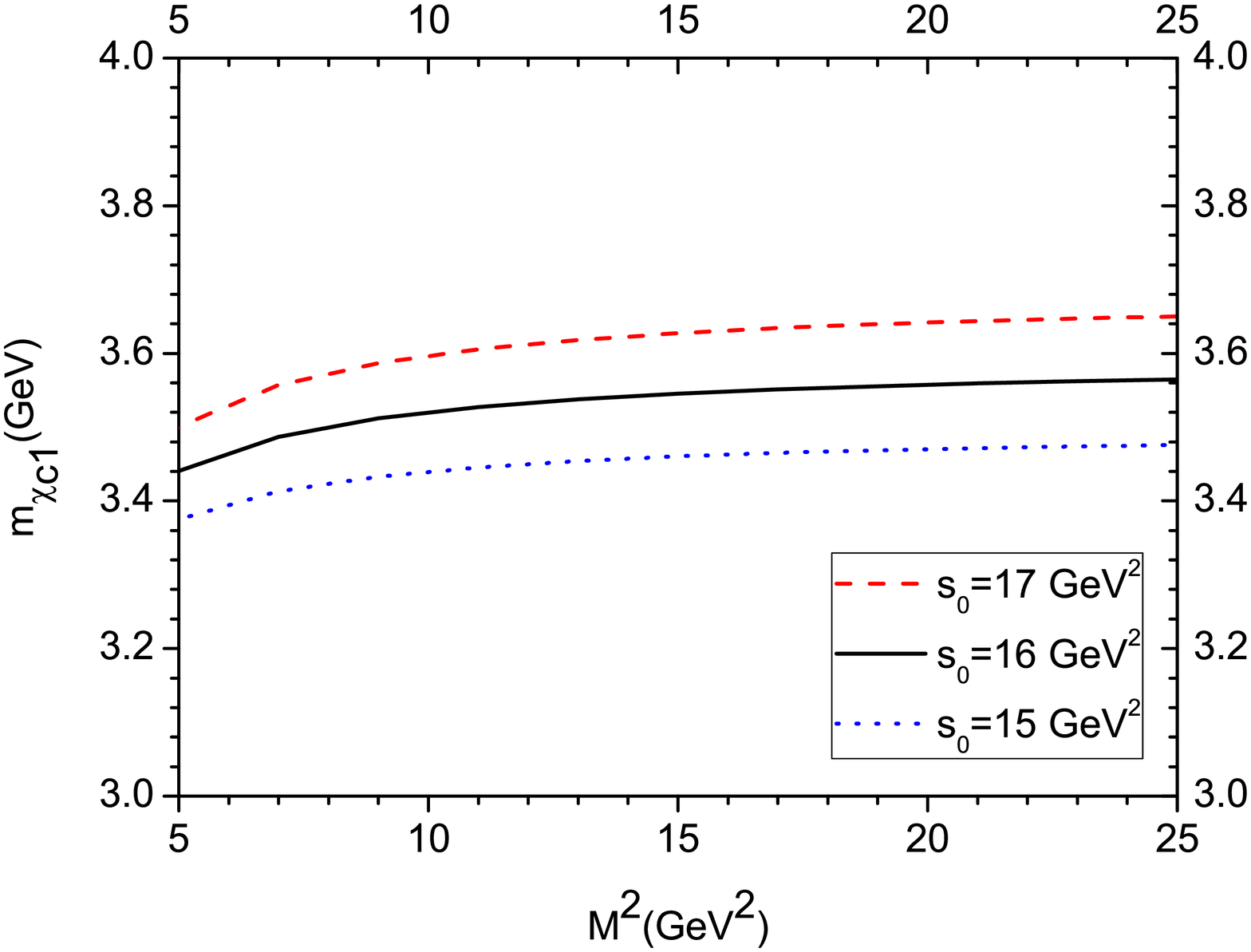}
\end{center}
\caption{Dependence of the mass of the $\chi_{c1}$ meson
on the Borel parameter $M^2$ at zero temperature.} \label{mYMsq}
\end{figure}
\begin{figure}[h!]
\begin{center}
\includegraphics[width=10cm]{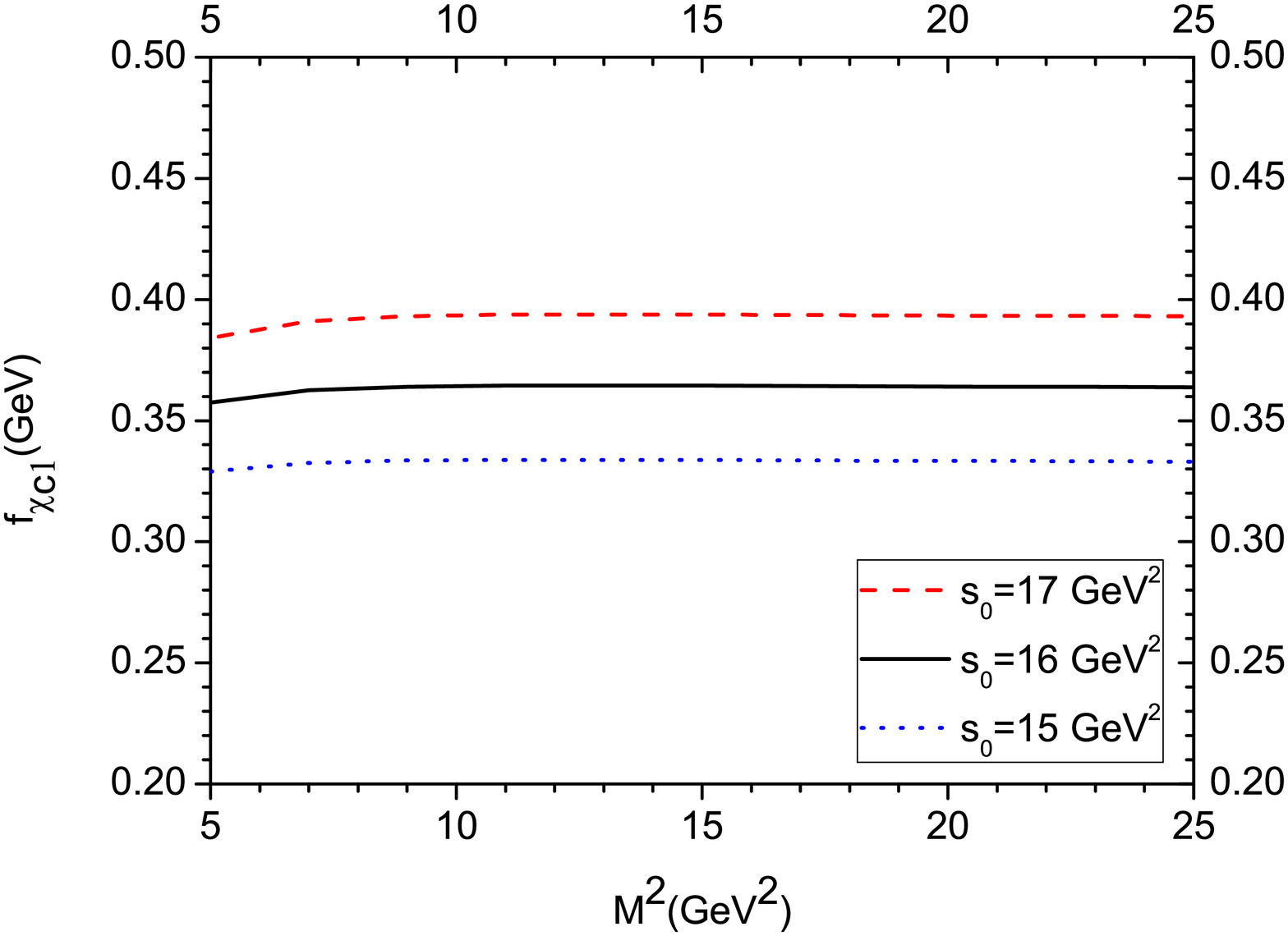}
\end{center}
\caption{Dependence of the  decay constant of the
$\chi_{c1}$ meson  on the Borel parameter $M^2$ at zero temperature.}
\label{fYMsq}
\end{figure}
\begin{figure}[h!]
\begin{center}
\includegraphics[width=12cm]{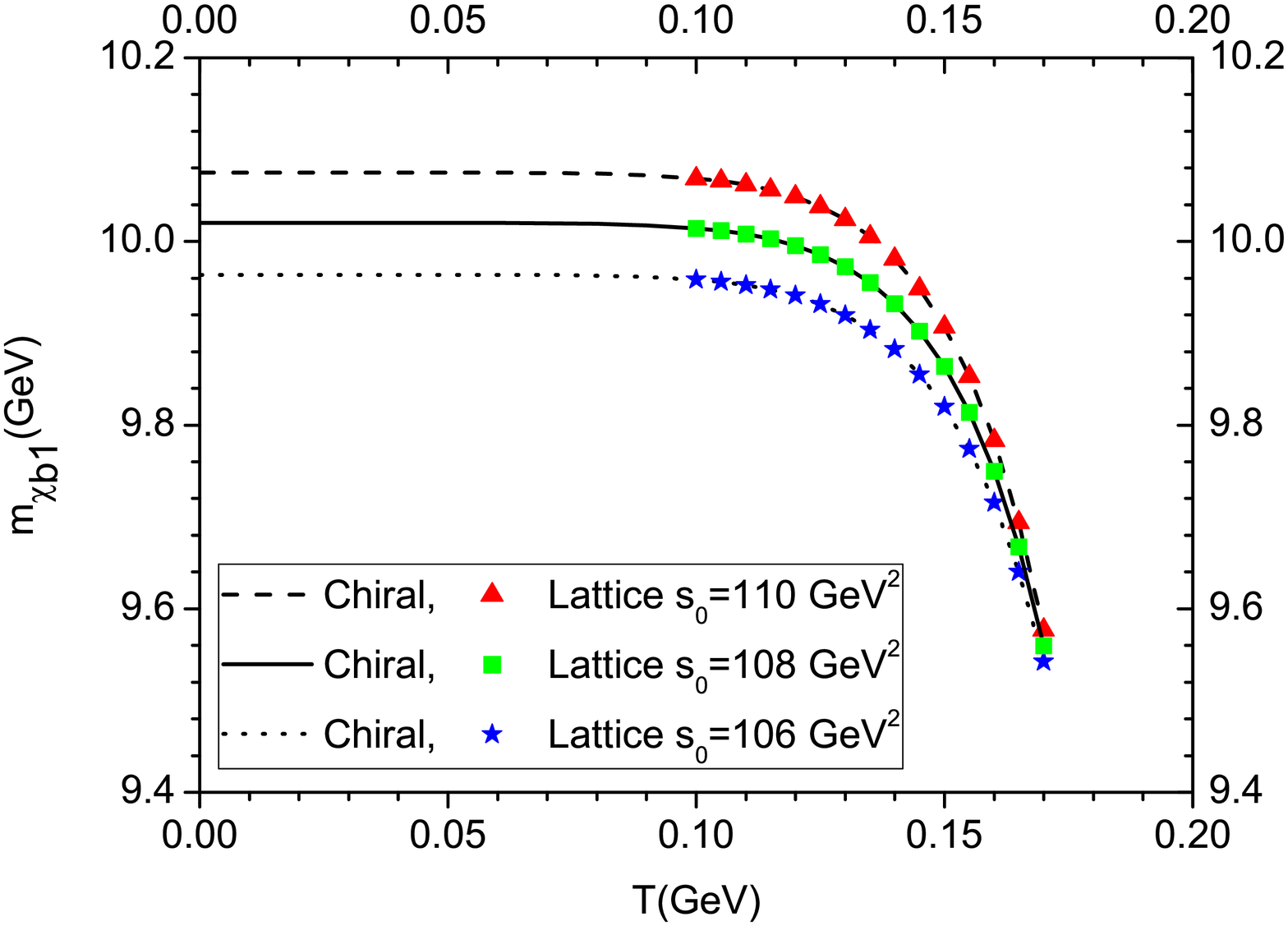}
\end{center}
\caption{Dependence of the mass of the $\chi_{b1}$  meson  on temperature at $M^2=20~GeV^2$.} \label{mJPsiTempMsq10Last}
\end{figure}
\begin{figure}[h!]
\begin{center}
\includegraphics[width=12cm]{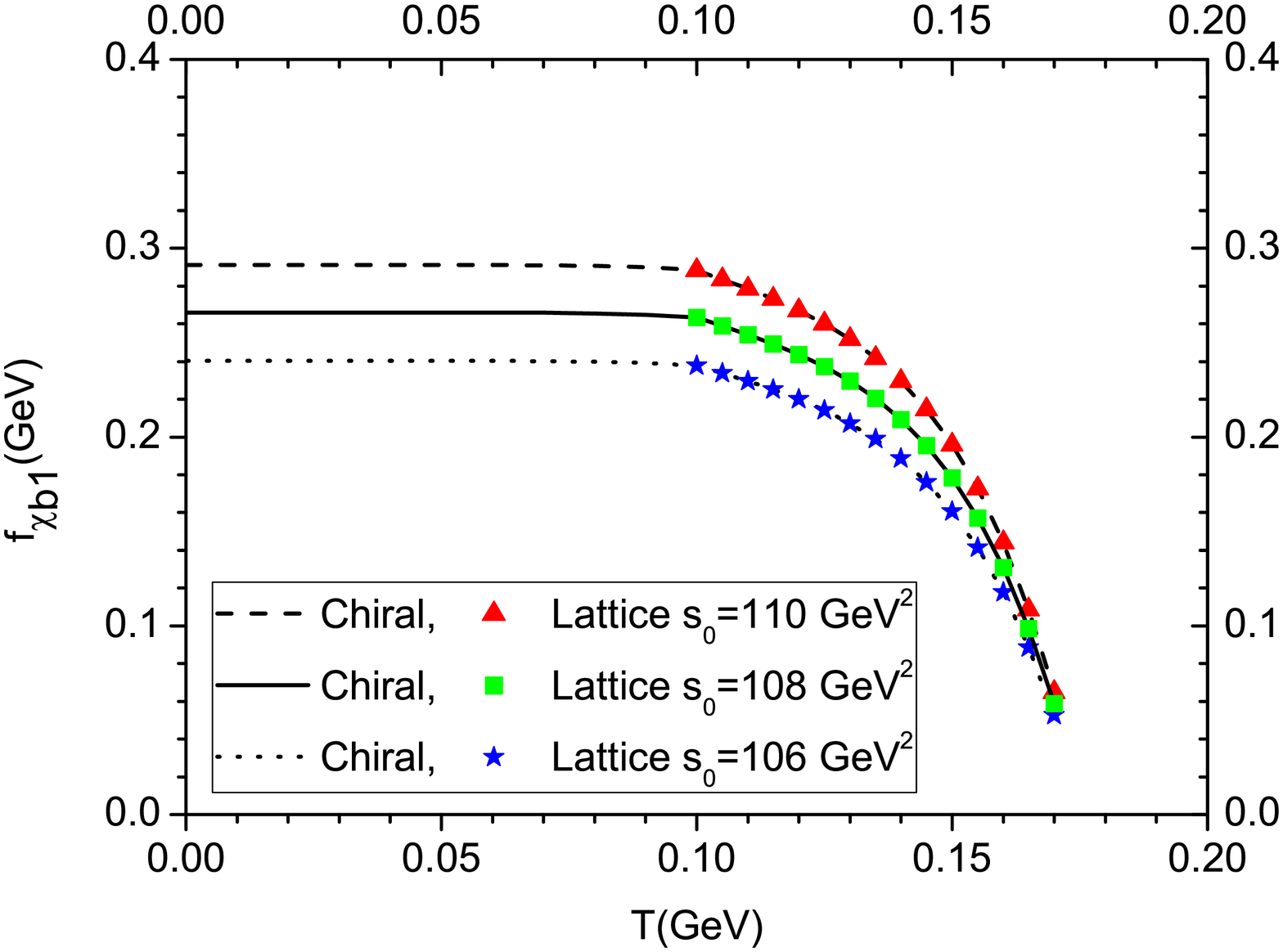}
\end{center}
\caption{Dependence of the  decay constant of the
$\chi_{b1}$  meson  on temperature at $M^2=20~GeV^2$.}
\label{fJPsiTempMsq10Last}
\end{figure}
\begin{figure}[h!]
\begin{center}
\includegraphics[width=12cm]{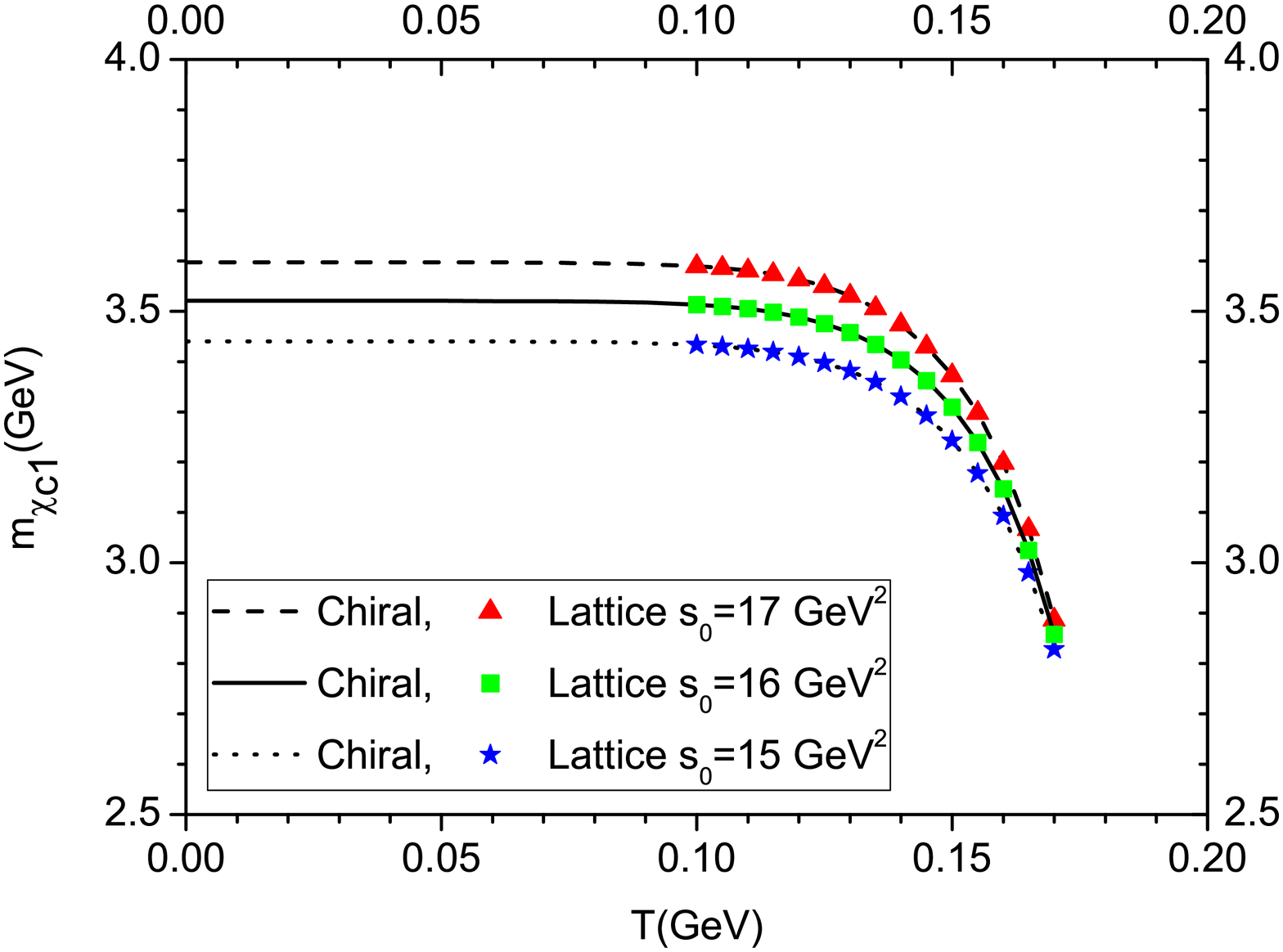}
\end{center}
\caption{Dependence of the mass of the $\chi_{c1}$  meson  on temperature at $M^2=10~GeV^2$.} \label{mYTempMsq20}
\end{figure}
\begin{figure}[h!]
\begin{center}
\includegraphics[width=12cm]{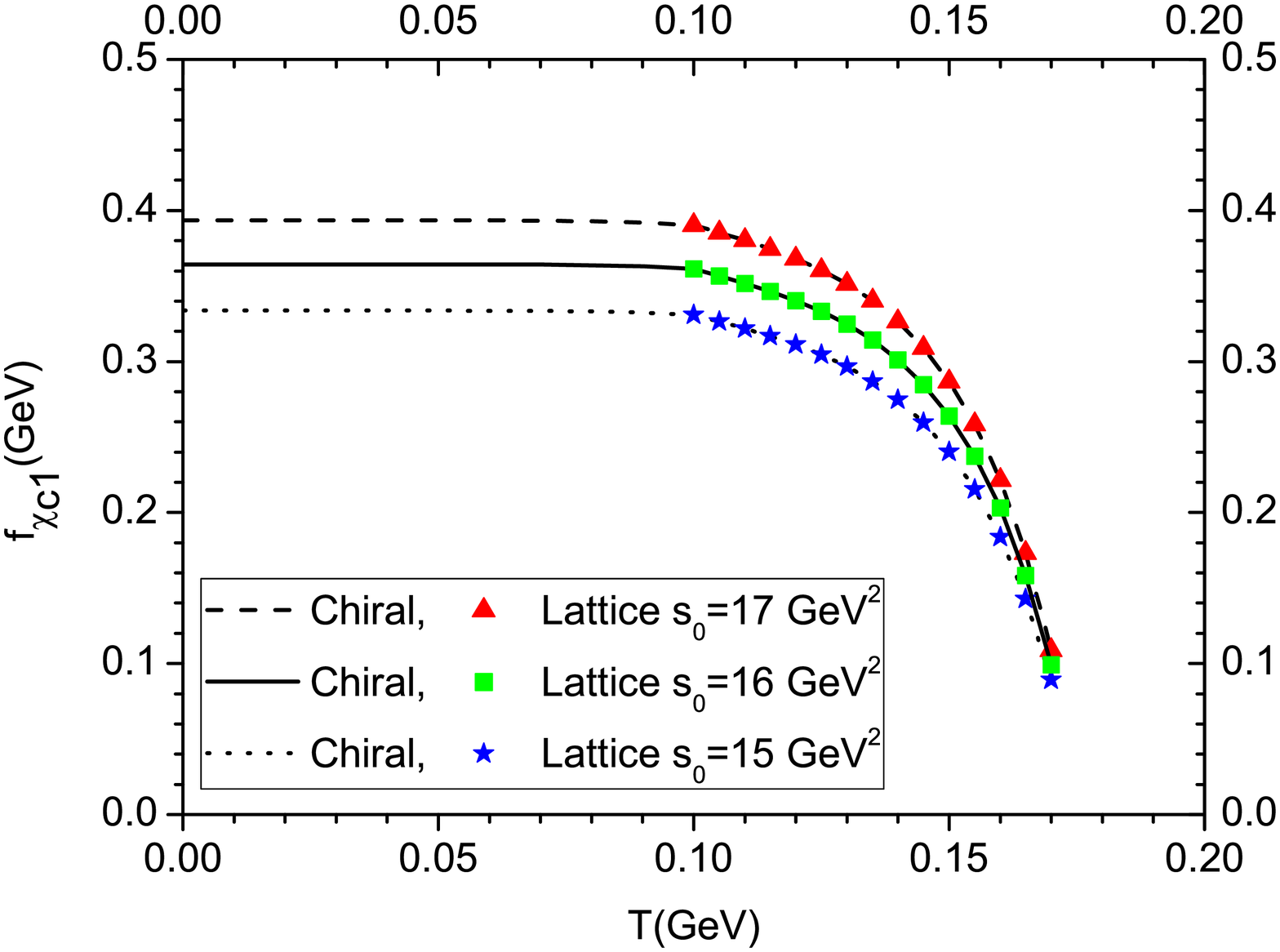}
\end{center}
\caption{Dependence of the decay constant of the
$\chi_{c1}$  meson  on temperature at $M^2=10~GeV^2$.}
\label{fYTempMsq20}
\end{figure}

Using the above obtained  working regions for auxiliary parameters
together with the other inputs, we plot the dependence on the Borel
parameter $M^{2}$ of the masses and
 decay constants of  the heavy axial  $\chi_{b1}$ and $\chi_{c1}$
 quarkonia  at zero temperature in Figs. (\ref{mJPsiMsq17Jan}-\ref{fYMsq}). From these figures, we see that the results weakly depend on the auxiliary parameters in their working regions.
The numerical results for the masses and decay constants of the heavy axial vector mesons under consideration are depicted in tables I and II. We also compare the obtained results with the experimental values in the
same tables. From table I we see a good consistency of our results with the experimental data. The errors in the results of our work belong to the uncertainties in calculation of the working regions for auxiliary parameters
as well as those coming from other inputs.

\begin{table}[h]
\renewcommand{\arraystretch}{1.5}
\addtolength{\arraycolsep}{3pt}
$$
\begin{array}{|c|c|c|c|}
\hline \hline
 &  m_{\chi_{c1}}~(GeV)& m_{\chi_{b1}}~(GeV)\\
\hline
  \mbox{Present Work }       &  3.52\pm0.11   &  9.96\pm0.26 \\
\hline
 \mbox{Experiment \cite{K. Nakamura}} &  3.51066\pm 0.00007 &9.89278\pm0.00026 \pm0.00031  \\
 \hline \hline
\end{array}
$$
\caption{Values for the masses of  the heavy axial  $\chi_{b1}$ and $\chi_{c1}$
 quarkonia  at zero temperature.} \label{tab:mass}
\renewcommand{\arraystretch}{1}
\addtolength{\arraycolsep}{-1.0pt}
\end{table}
\begin{table}[h]
\renewcommand{\arraystretch}{1.5}
\addtolength{\arraycolsep}{3pt}
$$
\begin{array}{|c|c|c|c|}
\hline \hline
         &f_{\chi_{c1}}(MeV) & f_{\chi_{b1}}(MeV)   \\
\hline
  \mbox{Present Work}        &  344\pm27   &  240\pm12 \\
                    \hline \hline
\end{array}
$$
\caption{Values for the decay constants of   the heavy axial  $\chi_{b1}$ and $\chi_{c1}$
 quarkonia  at zero temperature.} \label{tab:lepdecconst}
\renewcommand{\arraystretch}{1}
\addtolength{\arraycolsep}{-1.0pt}
\end{table}

At the end of this section we would like  to discuss the  behavior of the
 decay constants and masses of the heavy axial quarkonia under consideration in terms of temperature. We depict the variations of  these quantities versus temperature in
 figures (\ref{mJPsiTempMsq10Last}-\ref{fYTempMsq20}).  From these figures, we see    that the masses and
decay constants remain unchanged with the variation of
temperature up to $T\cong 100 ~MeV$. After this point they
start to decrease increasing the temperature. At deconfinement or
critical temperature, the decay constants decrease about (73-78)\%, while the masses are
decreased about 4\%, and 19\%  for $\chi_{b1}$ and $\chi_{c1}$
states, respectively. The sharp   decreasing in the values of the decay constants
near the deconfinement temperature can be considered as a signal for existing the QGP
as the new phase of  hadronic matter.

\section{Acknowledgement}
The authors are grateful to
 T. M. Aliev for useful
discussions. This work has been supported in part by the Scientific and
Technological Research Council of Turkey (TUBITAK) under the
research project No. 110T284 and research fund of Kocaeli University under grant No. 2011/029.

\end{document}